\documentclass[reqno,12pt]{article}
\usepackage{amsmath,amsfonts,amssymb,amsthm,amstext,amscd,eucal,xcolor}
\usepackage[all]{xy}
\usepackage{hyperref}
\usepackage{epsfig}
\usepackage{color}
\usepackage{graphicx}
\usepackage[active]{srcltx}
\makeatletter \@addtoreset{equation}{section}

\makeatletter\renewcommand\section{\@startsection {section}{1}{\z@}%
                                   {-3.5ex \@plus -1ex \@minus -.2ex}
                                   {2.3ex \@plus.2ex}%
                                   {\normalfont\large\bfseries}}
\renewcommand\subsection{\@startsection{subsection}{2}{\z@}%
                                     {-3.25ex\@plus -1ex \@minus -.2ex}%
                                     {1.5ex \@plus .2ex}%
                                     {\normalfont\bfseries}}

\newcommand{\be}{\begin{equation}}
\newcommand{\ee}{\end{equation}}
\newcommand{\bea}{\begin{eqnarray}}
\newcommand{\eea}{\end{eqnarray}}
\newcommand{\bse}{\begin{subequations}}
\newcommand{\ese}{\end{subequations}}
\newcommand{\bi}{\begin{itemize}}
\newcommand{\ei}{\end{itemize}}

\newcommand{\beq}{\begin{eqnarray}}
\newcommand{\eeq}{\end{eqnarray}}

\newcommand{\nn}{\nonumber}

\def\s2s1{S$^2\times$S$^1$ }
\def\Label#1{\label{#1}%
  \smash{\hbox to0pt{\raise1ex\hbox{\tiny[#1]}\hss}}}
\def\noLabels{\let\Label=\label}
\def\nobbibitem{\let\bbibitem=\bibitem}

\begin{document}
\baselineskip 18pt%
\begin{titlepage}
\hfill%
\hfill
\vbox{
    \halign{#\hfil         \cr
         IPM/P-2013/021  \cr
          } 
      }  
\vspace*{10mm}
\begin{center}
\centerline{{\Large{\textbf{Dual 2d CFT  Identification of Extremal Black Rings from  Holes}}}}
\vspace*{8mm}

{Ahmad Ghodsi\footnote{a-ghodsi@ferdowsi.um.ac.ir}$^{,a}$, Hanif Golchin\footnote{hanif.golchin@stu-mail.um.ac.ir}$^{,a}$, and M.M. Sheikh-Jabbari\footnote{jabbari@theory.ipm.ac.ir}$^{,b}$}\\
\vspace*{0.4cm}
{$^a$ \it Department of Physics, Ferdowsi University of Mashhad, \\
P.O.Box 1436, Mashhad, Iran}\\
{$^b$ \it School of Physics, Institute for Research in Fundamental Sciences (IPM),\\
 P.O.Box 19395-5531, Tehran, Iran}
\vspace*{1.0cm}
\end{center}

\begin{abstract}
Five dimensional Einstein gravity vacuum solutions in general fall into two classes of black rings with horizon topology $S^2\times S^1$, and black holes with horizon topology $S^3$. These solutions are specified by their mass and two spins.
There are ``overlapping'' regions of this parameter space where one has extremal rings and holes of the same spins. We show that for such regions the hole has
generically a larger entropy than the ring, and likewise, the central charge of the proposed chiral 2d CFT dual to the hole is larger than that of the
ring. For special places of this overlapping region where one of the spins tends to zero, the entropies of the extremal ring and hole also tend to zero and
essentially become equal. In this case we are dealing with Extremal Vanishing Horizon (EVH) black holes or rings. The near horizon geometry of the near-EVH hole
and rings both contain locally AdS$_3$ throats, providing a basis for the EVH/CFT proposal, a 2d CFT description of the low energy excitations of EVH hole or ring. We argue how the near-EVH hole and near-EVH ring can be distinguished from this dual 2d CFT viewpoint: The hole is a  thermal state with zero temperature in the left sector and finite temperature in the right, while the ring is a generic state in the ground state (of the CFT on the plane) in the left sector and a thermal state in the right.
The latter is part of the Hilbert space of the 2d CFT  obtained in the Discrete Light Cone Quantization (DLCQ).

\end{abstract}

\end{titlepage}
\tableofcontents
\section{Introduction}
A black hole is a stationary geometry which is generically specified by  a smooth, non-degenerate, codimension two  compact surface, the Killing horizon, the normal to which is a null Killing vector field. The Killing horizon depending on the gravity theory to which black hole is a solution, can have different topologies. In particular, for vacuum Einstein gravity with asymptotic flat black holes there are theorems which restrict horizon topology: In four dimensions we have the famous no hair theorem, stating that the horizon topology can only be $S^2$ \cite{HE}. In five dimension, the no-hair theorem does not hold; here  black hole solutions form a three parameter family specified by $(M,J_\phi,J_\psi)$, corresponding to mass and two spins. There are regions on this parameter space where we can have black hole solutions with $S^3$ horizon topology \cite{MP} as well as black ring solutions with $S^2\times S^1$ horizon topology \cite{Emparan:2001wn,Pomeransky:2006bd, Emparan:2008eg}. Explicitly, and as has been depicted in Figure \ref{fig2} and will be discussed in more detail in section \ref{section-4}, there is a region in the margins of the parameter space of black ring where ring and hole solutions overlap. This region denoted by the circle on the right-top of Figure \ref{fig2}, is where we can have extremal rings and hole with equal mass and spins.  This is the ``collapsing'' region \cite{Elvang:2007hs} in the ring/hole  parameter space.

In view of these possibilities one may ask several interesting questions, e.g. is it possible to have a (dynamical) topology change or if the hole and the ring can both form in a gravitational collapse process? One may then ask which one is the preferred outcome or which one is the more stable solution. Here we take on the latter question. As we will show for generic values of angular momenta in this overlapping region the black hole solution has a larger entropy than the ring (with the same angular momenta) and the hole is then expected to be a more stable configuration. There is, however, a specific limit in this ``collapsing region'' where the ring and the hole also have equal entropy, in the vanishing entropy (finite mass) limit. In this limit the entropy cannot be used to distinguish a ring from a hole. So the question is how one can differentiate between the two in the vanishing entropy limit from the near horizon data?

To address such questions one may use ``dual CFT'' description of  black holes, when applicable. For the class of asymptotic flat 5d vacuum Einstein gravity solutions which we will be considering in this paper, there are proposals for dual CFTs  for extremal black holes and rings \cite{Lu:2008jk, Chen:2012yd}.\footnote{In this work we will be dealing with extremal vacuum 5d Einstein solutions. Similar question, microstate counting of black rings and comparing it with black holes have also been discussed in the context of D1-D5-P system e.g. in \cite{D1D5P}.} For generic, finite entropy (extremal) case we show that the central charge of the chiral 2d CFT dual to the black hole is larger than that of the ring, while the vanishing entropy case is a bit different.

Dealing with vanishing entropy limit in the first sight may seem an obstacle in using the CFT dual description for the question of  ring/hole distinction. In fact,  quite on the contrary, being around an Extremal Vanishing Horizon (EVH) black hole/ring  provides a better ground for studying this problem. There is the EVH/CFT proposal \cite{SheikhJabbaria:2011gc} which associates a dual 2d CFT description for a generic near-EVH excitation of a given EVH black hole and this is to be contrasted with the Kerr/CFT proposal \cite{Guica:2008mu,Compere-review} according which the dual description of a generic extremal black hole/ring is a ``chiral 2d CFT''. The latter has much less dynamical content than a 2d CFT and may be related to it through the discrete light-cone quantization (DLCQ) \cite{Balasubramanian:2009bg}.

For several different examples of EVH black holes considered \cite{EVH-examples,deBoer:2011zt,Johnstone:2013eg}, although uniqueness theorems are still missing, it has been observed that one generically finds a locally AdS$_3$ space in the near horizon limit \cite{Johnstone:2013ioa}. Despite the fact that this AdS$_3$ happens to be a pinching orbifold of AdS$_3$ \cite{deBoer:2010ac}, it can be used to propose the EVH/CFT, a 2d CFT description dual to the near-EVH excitations \cite{SheikhJabbaria:2011gc,deBoer:2011zt,Johnstone:2013eg}. In this work we first review and discuss EVH black holes in the family of 5d Myers-Perry (MP) black holes and show that EVH MP black holes show a similar behavior. For near-EVH MP black holes this pinching AdS$_3$ turns into a pinching BTZ geometry. In particular, for extremal near-EVH MP black holes we find an extremal BTZ in the near horizon geometry.

Next, we consider  EVH black rings and their near horizon geometry. We show that the near horizon geometry of (near) EVH black ring again contains a locally AdS$_3$ space, a (pinching) self-dual AdS$_3$ orbifold \cite{deBoer:2010ac}. That is, for a given near-EVH geometry with the same spins and mass, the distinction between black ring or black hole can already be  distinguished from the near horizon geometry: the former has an extremal BTZ while the latter has a self-dual AdS$_3$ orbifold in the near horizon geometry.

We may then use this distinction to argue how the dual CFT can distinguish between a hole and a ring: As has been discussed in \cite{Balasubramanian:2009bg, deBoer:2010ac}, a generic extremal BTZ corresponds to a thermal state of the 2d CFT with vanishing temperature, say in a left moving sector, and a finite temperature in the right moving one. Whereas, the self-dual orbifold corresponds to a  non-thermal ground state in the left, and  a thermal state on right  sector \cite{Balasubramanian:2009bg}. The above fact about 2d CFTs leads to the proposal that the ring-hole transition is like a thermalization process in (the left moving sector of) the dual CFT. Here bring some preliminary evidence in support of this proposal, while it is desirable to explore it in further detail.

This paper is organized as follows. In section \ref{section-2}, we review and discuss 5d Myers-Perry (MP) black holes, their extremal and EVH limits and the corresponding near horizon geometries. In section \ref{section-3}, we review and discuss 5d black ``balanced'' rings and discuss in some detail the special places of their parameter space, in particular the near-EVH ring. In section \ref{section-4}, we review dual chiral 2d CFT description for generic extremal rings or holes and provide a dual 2d CFT description for the near-EVH black holes and near-EVH black rings. We discuss how the dual CFT can distinguish the ring from hole. In section \ref{section-5}, we make some concluding remarks and discuss some interesting open questions. In particular we briefly comment on the near-EVH ``unbalanced'' ring case and how it fits into our 2d CFT picture.

\section{Myers-Perry black holes and their EVH limit}\label{section-2}
In this section we review and study some facts about 5d Myers-Perry (MP) black holes. We focus on two special cases and their near horizon limits, the extremal MP which will be discussed in section \ref{section-2-1} and EVH and near-EVH MP, discussed in section \ref{section-2-2}.

In the most general form 5d MP black holes form a three parameter family of vacuum Einstein gravity solutions with metric \cite{MP, Hawking-Hartle}
\bea \label{MP}
ds^2\!\!\!\!&=&\!\!\!\! -\frac {\Delta}{{\rho}^{2}} \left( {dt}-a \sin^2\theta d\phi - \,b \cos^2 \theta d\psi \right) ^{2}+{\frac {{\rho}^{2}{{dr}}^{2}}{\Delta}}+{\rho}^{2}{d\theta }^{2}+ \frac {\sin^2 \theta}{{\rho}^{2}} \bigl( a\,{dt}- \left( {r}^{2}+{a}^{2}
 \right) d\phi  \bigr) ^{2} \nn \\ &+&\!\!\!\!\!{\frac { \cos^2 \! \theta}{{\rho}^{2}}\! \left( b\,{dt}-\! \left( {r}^{2}\!\!+\!
{b}^{2} \right)\! d\psi  \right) ^{2}}\!\!+\!{\frac {1}{{r}^{2}{\rho}^{2}}\! \left( ab\,{dt}-b\! \left( {r}^{2}\!\!+\!{a}^{2} \right) \sin^2\! \theta d\phi -a \left( {r}^{2}\!\!+\!{b}^{2} \right) \cos^2\! \theta d\psi  \right) ^{2}} \!,
\eea
where $\frac{3\pi M}{4G_5}$ is the (ADM) mass of the black hole, $a$ and $b$ are two rotation parameters and
\be \label{delt}
\Delta=\frac{1}{r^2}(r^2 + a^2)(r^2 + b^2) - 2M, \qquad \rho^2 = r^2 + a^2\cos^2\theta + b^2\sin^2\theta\,. \ee
The location of horizons can be found by solving  $\Delta=0$ equation; so the outer and inner horizons lie at
\be
r_\pm=\sqrt{M-\frac{a^2}{2}-\frac{b^2}{2}\pm \sqrt{ ( M-\frac12\, ( a-b ) ^{2} ) ( M-\frac12\, ( a+b ) ^{2} )}}\,.
\ee
The horizon (which is a Killing horizon at $r=r_+$) is parameterized by $\theta\in [0,\pi/2]$ and $\phi,\psi\in [0,2\pi]$ and its topology is $S^3$.

The Hawking temperature,  Bekenstein-Hawking entropy, angular momenta and  angular velocities at the horizon of this black hole are given by
\bea
T_H&=& \frac{r_+^4 - a^2 b^2}{2\pi r_+ (r_+^2 + a^2)(r_+^2 + b^2)}\,, \qquad \qquad S_{BH}= \frac{\pi^2 (r_+^2 + a^2) (r_+^2 + b^2)}{2\,G_5 r_+}\,, \label{MP-T-S}\\
J_{\phi}&=& \frac{\pi M a}{2G_5}\,,\qquad J_{\psi}= \frac{\pi M b}{2\,G_5}\,; \qquad \qquad \Omega_{\phi} = \frac{a}{r_+^2 + a^2}\,, \qquad \Omega_{\psi} = \frac{b}{r_+^2 + b^2}\,.\label{MP-Omega-J}
\eea
\subsection{Extremal MP black holes}\label{section-2-1}
Extremality occurs when $T_H=0$, i.e. when
\be\label{Ext-MP-parameters}
r_+^2=r_-^2=ab\,,\qquad M=\frac{1}{2} (a+ b)^2\,.
\ee
Extremal MP black holes, as depicted in Figure \ref{fig1}, are  specified by a paraboloid in $(M,a,b)$ space.
The Bekenstein-Hawking entropy for extremal black holes is then given as
\be\label{ext-MP-entropy}
S_{BH}= \frac{\pi^2}{2G_5} \sqrt{ab} (a+b)^2\,.\ee

The near horizon geometry of extremal MP black hole may be obtained through the following $\epsilon\to 0$ limit \cite{Bardeen:1999px, Matsuo:2010ut}
\be \label{nhtr}
t \to \frac{(a+b)^2}{4}\frac{\hat t}{\varepsilon}\,,\qquad r \to\sqrt{ab}+ \varepsilon \hat r\,,\qquad   \phi \to  \hat \phi+\frac{t}{a+b}\,,\qquad   \psi \to  \hat \psi+\frac{t}{a+b}\,,
\ee
while keeping the hatted coordinates fixed. Applying the above one finds
\bea \label{nhmp}
ds^2\!\!\!&=&\!\!\!
 \rho_0^2\big(- \frac{\hat r^2}{4} d\hat t^2
 + \frac{d\hat r^2}{4\hat r^2} \big)+ \rho_0^2d\theta^2 +\frac{2M}{\rho_0^2}\bigg[a^2\sin^2\theta
 \left( d\hat\phi + k_{\hat\phi}\hat r d\hat t
 \right)^2 \nn \\
 &+&\!\!\!b^2\cos^2\theta\! \left( d\hat\psi + k_{\hat\psi}\hat r d\hat t
 \right)^2\!\!+\! ab \left( \sin^2\,\theta ( d\hat\phi + k_{\hat\phi}\hat r d\hat t)
 +\cos^2\theta( d\hat\psi + k_{\hat\psi}\hat r d\hat t) \right)^2 \bigg]\,,
 \eea
 where
 \be
\rho^2_0 = ab +a^2\cos^2\theta + b^2\sin^2\theta , \qquad k_{\hat\phi} = \frac{1}{2}\sqrt{\frac{b}{a}},\qquad k_{\hat\psi}= \frac{1}{2}\sqrt{\frac{a}{b}}\,.
 \ee
As in general 5d near horizon extremal geometries \cite{Kunduri:2013gce}
, \eqref{nhmp} has $SL(2,R)\times U(1)_\phi\times U(1)_\psi$ isometry. Moreover, one can readily show that the volume of constant $r,t$ surfaces is equal to $4S_HG_5$.

\subsection{EVH MP black holes}\label{section-2-2}

Extremal Vanishing Horizon (EVH) black holes are in general defined as black holes with
\be\label{EVH-def}
T_H=0,\qquad A_H=0\,,
\ee
where $A_H$ is the horizon area and vanishing of $A_H$ and $T_H$ in the ``near-EVH'' region should happen such that \cite{SheikhJabbaria:2011gc}
\be\label{near-EVH-def}
T_H\to 0,\quad A_H\to 0\,,\quad A_H/T_H=finite.
\ee
Moreover, vanishing of horizon area should be a result of having a vanishing one-cycle on the horizon \cite{Johnstone:2013ioa}.
The solutions to \eqref{EVH-def}, if exist, in general define a codimension two (or in some cases larger) subspace in the parameter space.
Eq.\eqref{near-EVH-def} and the vanishing one-cycle condition on the horizon may then  be used to argue that in the near horizon limit of EVH black holes one expects to find an AdS$_3$, instead of the usual AdS$_2$ factor of extremal black holes \cite{Kunduri:2013gce}, where the AdS$_3$ is formed from  combination of the AdS$_2$ factor and the vanishing one-cycle which were on the horizon \cite{Johnstone:2013ioa}. This expectation has been checked for several EVH examples \cite{SheikhJabbaria:2011gc}, while still a general theorem on this is still missing.

\begin{figure}[ht]
\center
\includegraphics[width=100mm,height=140mm,angle=90]{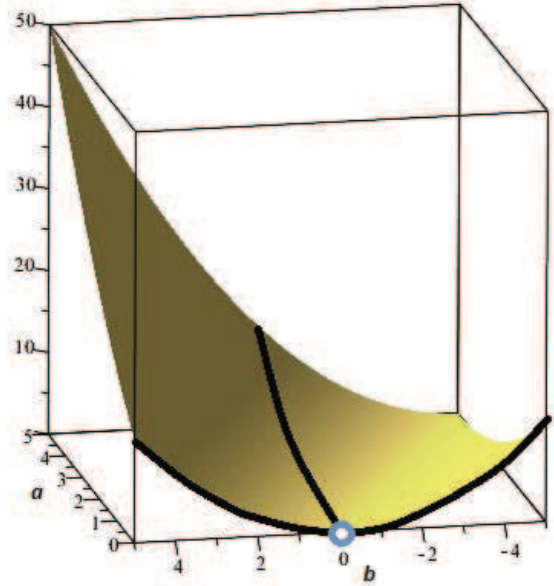}
\vspace{-5mm}
\caption{{\small The three dimensional parameter space of 5d MP black holes. Since only the relative sign of $a$ and $b$ parameters appears in the metric, we have depicted the parameter space in $a\geq 0$ region, while $b$ can be a generic real number. The $M=a=b=0$ is excluded because it does not correspond to a black hole (it is just 5d flat space). The two dimensional paraboloid surface corresponds to extremal black holes and the two thick black lines on it to EVH black holes. The extremal surface is a codimension one surface in this parameter space and the EVH lines, a codimension two surface.
}\label{fig1}}
\end{figure}

Let us now investigate the EVH black holes in the 5d MP family \cite{deBoer:2011zt}. Eq.\eqref{EVH-def} has the following solution
\be\label{EVH-MP-general}
M=\frac12(a+b)^2\,,\qquad a\,b=0\,.
\ee
Since we want to remain with a black hole solution of non-zero mass, the above has two solutions, $a=0,b\neq 0$ or $a\neq 0, b=0$, while the $a=b=0$ solution should be excluded. That is, single-spin 5d MP black holes are EVH black holes. The parameter space for these solutions has been depicted in Figure \ref{fig1}. Since the parameter space is symmetric with respect to $a\leftrightarrow b$ exchange, without loss of generality hereafter we restrict ourselves to $a=0$ case.

We now analyze the  near horizon geometry of an EVH (single rotation)  MP black hole (\ref{MP}) with $a=0$.  Let us start with the metric of the EVH MP black hole:
\bea
ds^2&=&-\frac{r^2-b^2\cos^2\theta}{r^2+b^2\sin^2\theta}dt^2+(r^2+b^2\sin^2\theta)(\frac{dr^2}{r^2}+d\theta^2)+r^2\sin^2\theta d\phi^2\cr &&\cr
&+&\frac{(r^2+b^2)^2-r^2b^2\cos^2\theta}{r^2+b^2\sin^2\theta}\cos^2\theta d\psi^2
-\frac{2b^3\cos^2\theta}{r^2+b^2\sin^2\theta} dt d\psi\,.
\eea
Performing the coordinate transformations
\be\label{EVHMP-NH}
\tilde r= r\, \epsilon\,,\qquad \tilde t= t/ \epsilon \,,\qquad \tilde\phi= \phi/\epsilon\,,\qquad  \tilde\psi \to  \psi+t/b\,,
\ee
and taking the limit $\varepsilon \to 0$, we obtain the following near horizon metric  \cite{Bardeen:1999px,deBoer:2011zt}
\be
ds^2=\sin^2 \theta \left(-\frac{\tilde r^2}{b^2}\,dt^2+ \frac{b^2}{\tilde r^2}\,d \tilde r^2+\tilde r^2 d \tilde \phi^2 \right)+ b^2 (\sin^2\theta\, d \theta^2+\cot^2\theta\, d\tilde \psi^2)\,.
\ee
One should note that, as in other EVH cases studied in the literature, the $AdS_3$ factor appearing in the near horizon geometry is a \textit{pinching} AdS$_3$ \cite{SheikhJabbaria:2011gc}, that is $\tilde\phi\in [0,2\pi \epsilon]$.
Using a reduction ansatz as \cite{deBoer:2011zt}
\be
ds^2=\sin^2\theta g_{\mu \nu} dx^{\mu} dx^{\nu}+ b^2 (\sin^2\theta\, d \theta^2+\cot^2\theta\, d\tilde \psi^2)\,,
\ee
the 5d Einstein-Hilbert action reduces to a 3d AdS$_3$ Einstein gravity with AdS$_3$ radius $\ell$ and 3d Newton constant $G_3$:
\be\label{3d-reduction-l-G}
\ell=b\,,\qquad G_3=\frac{G_5}{\pi b^2}\,.
\ee

\subsection{The near-EVH limit}\label{near-EVH-hole-section}
As depicted in Figure \ref{fig1}, one can move away from the EVH line in two directions transverse to the line. Again as is seen from the figure, one may move away from the EVH line while still remaining inside the extremal surface, or can move transverse to it, leaving the extremal surface. That is,  near-EVH ``excitations'' can be extremal or non-extremal. The allowed region for near-EVH parameter space is the one satisfying \eqref{near-EVH-def}. To see where on $(M,a,b)$ space it corresponds to,
following \cite{Johnstone:2013eg}, let us parameterize the near-EVH deviations as
\be \label{cov}
a=\delta a\,, \qquad b=b_0+\delta b\,, \qquad M=\frac{b_0^2}{2}+(\delta M+b_0\delta b)\,.
\ee
$T_H\sim S_{BH}\sim \epsilon$ is achieved if $r_\pm\sim \epsilon$ implying that $\delta a\sim \epsilon^2\,,\  \delta M\sim \epsilon^2$ so we choose
\be
\delta M=m\epsilon^2\,,\qquad a=\hat a\epsilon^2\,.
\ee
(Recalling \eqref{cov} and the extremality bound for MP black hole we learn that $m\geq 0$ and in our conventions we have chosen $\hat a\geq 0$.)

We now combine the above near-EVH parametrization with near horizon limit scalings of \eqref{EVHMP-NH} to obtain the near horizon near-EVH geometry. We choose $\epsilon$ to   parameterize both of the deviations from EVH and the near horizon.
In the near horizon limit $r=\hat r\, \epsilon$, the above scaling leads to
\be
\Delta=\frac{\epsilon^2}{\hat r^2 \ell^2}\left(\hat r^4-2\,\hat r^2 m+b_0^2\, \hat a^2 \right)=\epsilon^2 \hat \Delta\,,
\ee
where $\ell=b_0$ is the AdS$_3$ radius.
The near horizon of the near-EVH solution can be obtained by inserting
\be
r=\hat r\, \epsilon\,,\qquad t=\hat t/ \epsilon \,,\qquad \phi=\hat \phi/\epsilon\,,\qquad \Delta=\epsilon^2 \hat \Delta\,, \qquad \psi \to \hat \psi+t/b\,,
\ee
and (\ref{cov}) in the metric (\ref{MP}) and taking $\epsilon \to 0$  limit. The result is
\be\label{near-EVH-MP}
ds^2=\sin^2 \theta \left[-\hat \Delta\, d\hat t^2+\frac{d\hat r^2}{\hat \Delta}  + \hat r^2 \big(d\hat \phi-\frac{\hat a}{\hat r^2}d\hat t \big)^2 \right]+\ell^2 (\sin^2\theta\, d \theta^2+\cot^2\theta\, d\hat \psi^2)\,.
\ee
As we see in the near-EVH case the pinching AdS$_3$ part of the near horizon geometry has turned into a pinching BTZ black hole. If we ignore the pinching for the moment and use the usual formulas for mass and angular momentum of the BTZ, they are\footnote{For vanishing  $\hat a$ and $ m$ we  remain on the EVH line and it does not correspond to a deviation from EVH. This is  made explicit  noting that neither of the charges of the near horizon BTZ geometry do not depend on $\delta b$. }
\be
M_{BTZ}=\frac{1}{4\ell^2 G_3}\,\frac{\hat r^2_+ +\hat r^2_-}{2}=\frac{\pi m}{4 G_5 }\,,\,\qquad J_{BTZ}=\frac{r_+ r_-}{4\ell G_3}=\frac{\pi \hat a b_0^2}{4 G_5}\,,
\ee
where we have used the fact that $\hat r_\pm$ are roots of $\hat\Delta=0$ and \eqref{3d-reduction-l-G}. One may also compare the temperature and entropy of the BTZ black hole obtained in the near horizon limit and those of the original near-EVH MP black hole:
$$
T_{MP}=\frac{\hat r_+^2-\hat r_-^2}{2\pi \hat r_+ \ell^2}=T_{BTZ}\epsilon\,,\qquad S_{MP}= \frac{\pi^2 \hat r_+ b_0^2}{2 G_5}\epsilon= \frac{(2\pi\epsilon) \hat r_+}{4 G_3}=S_{BTZ}\,.
$$
One can also show that $\Omega_\phi=\frac{\hat a}{\hat r_+^2}=\frac{\hat r_-}{\ell r_+}=\Omega_{BTZ}$ where $\Omega_{BTZ}$ is the horizon angular velocity of the BTZ black hole and that
$$J_{BTZ}\epsilon^2=J_{\phi}\,,\qquad 2M_{BTZ}\epsilon^2=\frac{\pi M}{2G_5}-\Omega_\psi J_\psi+\Omega_\phi J_\phi\,.$$
This is in line with the discussions of \cite{Johnstone:2013eg,Johnstone:2013ioa}.

\section{Vanishing horizon limit of (balanced) rotating black ring}\label{section-3}
In this section we investigate the EVH conditions for neutral double rotating black ring (DRBR) \cite{Pomeransky:2006bd} and discuss its near-EVH limit.
DRBR is a stationary solution of 5d vacuum Einstein gravity with horizon topology  $S^1 \times S^2$ and admits extremal limit. In the following, along with most of the black ring literature in particular \cite{Emparan:2001wn,Emparan:2008eg}, unlike \cite{Pomeransky:2006bd}, we use $\psi$ to denote the ring direction of and $\phi$ parameterizes the angle in the $S^2$. The metric is
\bea \label{metr1}
ds^2&=& -\frac{H(\hat y,x)}{H(x,\hat y)}(d\hat t\!+\!\Omega(x,\hat y))^2\!-\!\frac{F(x,\hat y)}{H(\hat y,x)}d\hat \psi^2-2\frac{J(x,\hat y)}{H(\hat y,x)} d\hat \phi\, d\hat \psi
\nn \\ &+&\frac{F(\hat y,x)}{H(\hat y,x)}d\hat \phi^2+\frac{2 k^2 H(x,\hat y)}{(x\!-\!\hat y)^2(1\!-\!\nu)^2}\big(\frac{dx^2}{G(x)}\!-\!\frac{d\hat y^2}{G(\hat y)}\big)\,,
 \eea
 where the range of coordinates are restricted to, $-1\leq x \leq 1$, $-\infty < \hat y <-1$ and $\phi,\psi \in [0,2\pi]$. The functions in the above metric are defined as follows
\bea\label{omega}
G(x)\!\!\!\!&=&\!\!\!\!(1\!-\!x^2)(1\!+\!\lambda x\!+\!\nu x^2)\,,\nn \\
 H(x,\hat y)\!\!\!\!&=&\!\!\!\! 1\!+\!\lambda ^2\!-\!\nu^2\!+\!2\lambda\nu (1\!-\!x^2)\hat y\!+\!2x\lambda(1\!-\!\hat y^2\nu^2)\!+\! x^2 \hat y^2 \nu(1\!-\!\lambda^2\!-\!\nu^2)\,,\nonumber\\
\Omega(x,\hat y)\!\!\!\!&=&\!\!\!\!\frac{\!-\!2 k \lambda \big((1+\nu )^2\!-\!\lambda ^2\big)^\frac12}{H(\hat y,x)} \big(\nu^\frac12 \hat y(1-x^2)d\phi\nn \\ &+&\frac{1\!+\!\hat y}{1\!-\!\lambda \!+\!\nu}(1+\lambda -\nu +\nu (1-\lambda-\nu)\hat yx^2+2\nu x(1-\hat y))d\psi),\nn\\
J(x,\hat y)\!\!\!\!&=&\!\!\!\!\frac{2 k^2 (1-x^2) (1-\hat y^2) \lambda  {\nu^\frac12}}{(x-\hat y) (1-\nu)^2} \,(1+\lambda ^2 -\nu ^2 + 2 (x+\hat y) \lambda  \nu-x \hat y \nu (1-\lambda ^2-\nu^2))\,, \nonumber\\
F(x,\hat y)\!\!\!\!&=&\!\!\!\! \frac{2 k^2}{(x\!-\!\hat y)^2 (1\!-\!\nu)^2}\! \Big(\!G(x) (1\!-\!\hat y^2)\big(((1\!-\!\nu)^2\!-\!\lambda ^2)
(1\!+\!\nu )\!+\!\hat y \lambda (1\!\!-\!\!\lambda ^2\!+\!2 \nu \!-\!3 \nu ^2)\big)\nn \\ &+& G(\hat y) (2 \lambda ^2 \!+\! x \lambda ((1\!-\!\nu )^2\!+\!\lambda ^2)+ x^2\big((1\!-\!\nu )^2\!-\!\lambda ^2\big) (1\!+\!\nu)\nn \\ &+& x^3\lambda(1\!-\!\lambda^2\!-\!3\nu^2\!+\!2\nu^3)\!-\! x^4 (1\!-\!\nu ) \nu (\lambda ^2\!+\!\nu ^2\!-\!1))\Big)\,.
\eea
The above constitutes a three parameter family of solution, parameterized by $k$, $\lambda$ and $\nu$; the latter two are dimensionless and $k$ is of dimension of length and is proportional to the radius of the ring direction. Ranges of these parameters are
\be \label{ranges}
k>0\,, \quad \qquad 0\leq\nu<1\,, \quad \qquad  2\sqrt{\nu}\leq\lambda<1+\nu\,.
\ee
Horizon (larger root of $G(y)$) is located at \be
y_{h}=\frac{-\lambda + \sqrt{\lambda^2-4\nu}}{2\nu}\,.
\ee
The entropy, Hawking temperature, angular velocities, angular momenta  and mass of this solution are given by \cite{Elvang:2007hs}
\bea \label{2scharges}
S_{BH}\!&=&\!\!\frac{A}{4G_5}=\frac{8 \pi^2 k^3 \, \lambda (1+\nu+\lambda)}{G_5(1-\nu)^2(y_h^{-1}-y_h)} \,,\qquad \quad T_{H}=\frac{(y_h^{-1} - y_h) (1-\nu) \sqrt{\lambda^2 - 4 \nu}}{8\pi\, k\, \lambda (1+\nu +\lambda)} \,,\nn \\
 J_{\psi}&=&\frac {2\,\pi{k}^{3}\lambda \sqrt{ \left( 1+\nu \right) ^{2}-{\lambda}^{2}} \left( {\nu}^{2}\!+ \left( \lambda-6 \right) \nu+\lambda+1 \right) }{G_5 \left( 1-\nu \right) ^{2} \left( 1+\nu-\lambda \right) ^{2}}\,,\qquad \Omega_{\psi}=\frac{1}{2 k} \sqrt{\frac{1+\nu-\lambda}{1+\nu+\lambda}} \,, \nn\\
J_{\phi}&=&\frac {4\,\pi \,\lambda \sqrt{\nu}\,{k}^{3} \sqrt{ \left(1+\nu \right) ^{2}\!-\!{\lambda}^{2}}}{G_5 \left( 1-\nu \right) ^{2} \left(1+\nu-\lambda \right) }\,, \qquad \Omega_{\phi}=\frac{\lambda (1+\nu)-(1-\nu)\sqrt{\lambda^2 - 4\nu}}
 {4 k\, \lambda \sqrt{\nu}} \sqrt{\frac{1+\nu-\lambda}{1+\nu+\lambda}}\,,
  \nn\\
M&=&\frac32\,(T_H S_{BH} + J_\phi \Omega_\phi + \Omega_\psi J_\psi) =\frac {3\,\pi \,{k}^{2}\lambda}{G_5(1+\nu-\lambda)}\,.
\eea

\subsection{Parameter space of the ring}\label{ring-parameter-space-section}
\begin{figure}[ht]
\center
\includegraphics[width=90mm,height=120mm,angle=90]{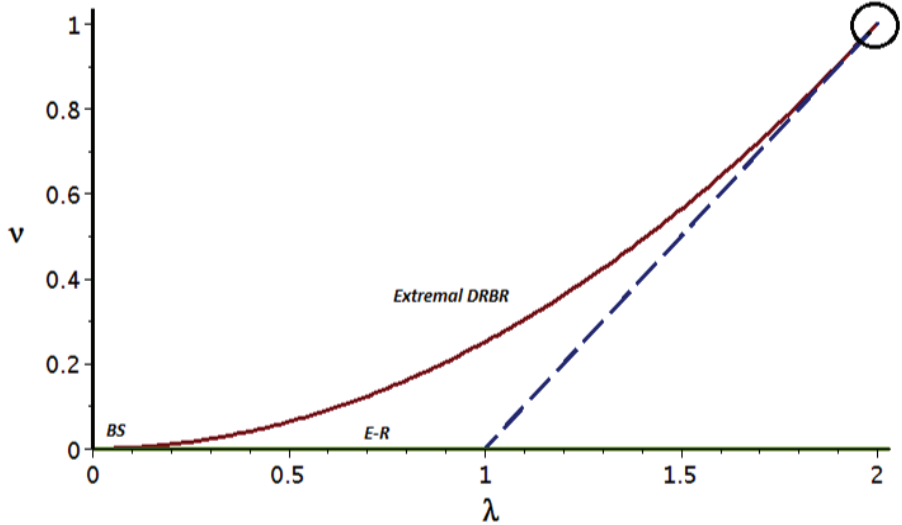}\vspace{-10mm}
\caption{{\small Parameter space of doubly rotating ring (the $k$ direction is suppressed). Extremal solutions lies on the parabola; the dashed line is not in the parameter space of rings;  the line $\nu=0$ present Emparan-Reall singly rotating black ring (E-R) and $\lambda\,,\nu\to0$ corner is the region of boosted Kerr strings (BS).}\label{fig2}}
\end{figure}

Similar to the case of MP black hole, there is a three dimensional parameter space for DRBR characterized by ($k\,, \lambda\,, \nu$). Ignoring $k$ direction, this parameter space is depicted in Figure \ref{fig2}.
Each point inside this figure corresponds to a {generic} black ring with three independent conserved charges, two spins and the mass. We now concentrate on the boundaries of the parameter space. For a thorough analysis of possible interesting curves inside the allowed region see \cite{Emparan:2008eg,Elvang:2007hs}.

\paragraph{\bf{Extremal ring, $\nu=\frac{\lambda^2}{4}$ curve}.}
{} From (\ref{2scharges}) we see that temperature vanishes on this curve, so this is the location of the extremal ring. The near horizon geometry of the extremal black rings can be obtained using the coordinates transformation \cite{Chen:2012yd, cringholography}
\bea \label{nhdrbr}
\hat y\! \to\frac{-2}{\lambda}\!+\epsilon y, \quad \hat t\to\! \frac{16\,k}{(\lambda\!-2)^2}\,\frac{t}{\epsilon},\quad \hat \phi \!\to\! \phi\!+\!\frac{(\lambda\!-2)(\lambda^2\!+4)}{8\,\lambda\,k\,(\lambda+2)}\,\hat t, \quad \hat \psi \!\to\! \psi\!+\!\frac{(\lambda-2)}{2k(\lambda+2)}\,\hat t\,,
\eea
 and taking the  $\epsilon \to 0$ limit. The result is
\bea \label{dringNH}
ds^2&=&{\frac {16k^2\Gamma(x)}{ ( \lambda-2 ) ^2}}\big(-y^2 dt^2+\frac{dy^2}{y^2} \big) +{\frac{8 {\lambda}^{2}{k}^{2}H ( x ) }{ ( \lambda x+2 ) ^{4} ( 1-{x}^{2} )  ( 4-{\lambda}^{2} ) }}{{ dx}}^{2}\\
 &+&4{\frac {{k}^{2} ( 2+\lambda ) ^{2}}{ ( 2-\lambda ) ^{2}}{d\psi }^{2}}+\frac{32{\lambda}^{2}{k}^{2} ( 1-x^2 )}{   H(x)( 4-\lambda^2 )} \big( d \phi -y{dt}+{\frac {( 4+8\lambda+\lambda^{2} )}{4\lambda}d\psi}\big)^{2}\,, \nn \\ H(x)\!\!&=&\!\!\left( {\lambda}^{2}+4 \right)  \left( 1+{x}^{2} \right) +8\,\lambda\,x\,,\qquad \quad \Gamma(x)=\frac{{\lambda}^{2}H(x)}{2 \left( 2+\lambda\,x \right) ^{2}
 \left( 2+\lambda \right) ^{2}}\,.
 \eea
It is interesting to note that the ring direction $\psi$ has a constant radius and the geometry takes form of the 4d metric times $S^1$. To see this more clearly, one may perform the coordinate change from $\phi$ to $\varphi=\phi+\frac {( 4+8\lambda+\lambda^{2} )}{4\lambda}\psi$, while the circle $S^1$ is parameterized by $\psi$.
This is the same geometry that one would find in the near horizon limit of an extremal boosted Kerr string \cite{cringholography}. One may readily check that the area of constant $y,\ t$  surface (where the horizon of the original extremal black ring was located) is
\be\label{horizon-area-extremal-ring}
A_H=\frac{32 k^3\lambda^2}{(2-\lambda)^2} \cdot (2\pi)^2 \int_{-1}^1 \frac{dx}{(2+\lambda x)^2}= \frac{64 \lambda^2}{2+\lambda}\frac{k^3}{(2-\lambda)^3} \cdot (2\pi)^2\,.
\ee

\paragraph{{\bf The $\lambda=1+\nu$ line.}} This line is not formally in the parameter space of the ring, nonetheless, one may study $\lambda\to1+\nu $ limit. Here we briefly mention the behavior of charges and thermodynamical parameters. One can readily check using \eqref{2scharges} that in the limit $1+\nu-\lambda=\hat \lambda \epsilon^2$ and $k=\hat k \epsilon\ \ , \epsilon\to 0$,  the mass $M$ and $J_\psi$ remain finite while $J_\phi\sim \epsilon^2\to 0$. The angular velocities $\Omega_\phi$ and $\Omega_\psi$ also remain finite. The entropy $S$ and temperature $T$ both go to zero (as $\epsilon$) such that $S/T$ remains finite (and $J_\phi\sim T_H^2\to 0$). This limit seems to be an interesting one since it formally falls into our definition of EVH rings. We also note that except for the $\nu=0$ case where we recover the single rotating ER ring \cite{Emparan:2001wn}, the $\lambda=1+\nu$ does not correspond to a ring. We will make more comments on this in the discussion section.

\paragraph{{\bf The $\nu=0$ line.}} On this line the $\phi$-angular momentum $J_\phi$ vanishes while the other charges and thermodynamical parameters remain finite. On this line we hence recover single-rotating ring of  Emparan-Reall  \cite{Emparan:2001wn}.

\paragraph{{\bf The $\lambda\,,\nu \to 0$ corner.}}
In the $\lambda=\nu=0$ point the ring degenerates and one can show that there is a particular $\lambda,\nu\to 0$ limit where the ring degenerates to boosted Kerr string solution \cite{Matsumoto:2012cg}. Explicitly, consider the following scaling
\be
\nu=\frac{a_*^2 \lambda^2}{4}\,,\quad \hat M=\frac{\lambda k}{\sqrt{2}}\,,\qquad \nu\to 0\,,\qquad 0\leq a_*\leq 1\,,
\ee
while keeping $\hat M, a_*$ fixed. (The range for $a_*$ is deduced from \eqref{ranges}.) By inserting the above scaling together with coordinate scaling \cite{Matsumoto:2012cg}
\be
\hat y=-\frac{\sqrt{2}\,k}{r}\,, \qquad x=\cos \theta\,, \qquad \hat \psi=-\frac{z}{\sqrt{2}\,k}\,,
\ee
into the metric of DRBR, one can obtain boosted string solutions as
\bea \label{BKBS}
ds^2\!\!\!\!\!&=&\!\!\!\!-\big(1\!-\!\frac{2\hat Mr\!\cosh^2\sigma}{\rho^2}\big)dt^2\!+\frac{2\hat Mr\!\sinh2\sigma}{\rho^2}\,dtdz \!+\!\big(1\!+\!\frac{2\hat Mr\sinh^2\sigma}{\rho^2}\big)dz^2\!+\frac{\rho^2}{\Delta}dr^2\!+\rho^2d\theta^2 \nn \\
\!\!\!\!\!&+&\!\!\!\!\!\frac{\left(r^2\!\!+\!a^2\right)^2\!-\!\Delta a^2\sin^2\!\theta}{\rho^2}\sin^2\!\theta d\phi^2 \!-\!\frac{4\hat Mr\!\cosh\!\sigma}{\rho^2}a\sin^2\!\theta dtd\phi\!-\!\frac{4\hat Mr\!\sinh\!\sigma}{\rho^2}a\sin^2\!\theta dzd\phi,
\eea
where $\sigma$ is boost parameter, $a=\hat M a_*$ is rotation parameter of Kerr and   $\rho^2=r^2+a^2 \cos^2 \theta,\, \Delta=r^2-2\hat Mr+a^2$. The special cases of $a_*=0$, $a_*=1$ respectively correspond to  boosted Schwarzschild string and extremal boosted Kerr string.

\paragraph{\bf {The ``collapsing point'' $\lambda=2, \nu=1$ \cite{Elvang:2007hs}.}} This is the upper end point on the extremal DRBR and is not strictly speaking in the parameter space of the ring (\emph{cf.}\eqref{ranges}). This is a singular point at which the entropy vanishes, so in our terminology, if the condition \eqref{near-EVH-def} is also satisfied, this can be an EVH point. At this point one of the angular momenta $J_\phi$ vanishes and the ring becomes an extremal single rotating ring, which according to the analysis of \cite{Emparan:2001wn} does not exist; this is ``collapsing point'' in the parameter space of the rings \cite{Elvang:2007hs}. Nonetheless, one may study approaching this point from inside the DRBR allowed region. Recalling \eqref{ranges} and demanding the mass $M$ to remain finite one can take the following limit:
\be\label{nu-lambda-range}
\nu=1-\rho\,,\qquad \lambda=2-\rho-\delta\,,\qquad 0<\delta\leq \rho^2/4\,,\qquad \rho,\delta\to 0\,,
\ee
or more explicitly, we keep $\sigma, \tilde M$,
\be \label{cop}
\sigma=\frac{1+\nu-\lambda}{(1-\nu)^2}\,,\qquad \tilde M=\frac{8k^2}{1+\nu-\lambda}\,,
\ee
fixed as we take $\nu\to 1,\lambda\to 2$ limit. Note that $0<\sigma\leq 1/4$.

It has been shown in \cite{Elvang:2007hs} that if we focus onto to a region on the geometry close to $x=-1,\ y=-1$ we find an \textit{extremal} MP black hole. Explicitly, consider
\be \label{transj}
x=-1+\frac{16 \sqrt{a} k^3 \cos^2\theta}{(a+b)^{3/2}(r^2-ab)}\,, \quad \quad \hat y=-1-\frac{16 \sqrt{a} k^3 \sin^2\theta}{(a+b)^{3/2}(r^2-ab)}\,,
\ee
The rotation parameters $a,b$ are
\be \label{cop2}
a=\sqrt{2\tilde M} \sigma\,,\qquad b=\sqrt{2\tilde M}(1-\sigma)\,.
\ee
Next recall that with the above expansion the horizon is located at
\be
y_h\simeq -1-\sigma (1-\nu)\,.
\ee
Therefore, with \eqref{transj} we are cutting the geometry close to its horizon, while \eqref{transj} is not a near horizon limit. The coordinate transformation and limit \eqref{transj} also cuts a part of the ``collapsing'' horizon of the ring and completes it such that its topology becomes $S^3$. In this region the metric takes exactly the form of an extremal MP black hole with the above mentioned parameters \cite{Elvang:2007hs}. This transformation, as discussed in \cite{Elvang:2007hs} is designed such that it maps an extremal ring of mass $M$ and angular momenta $J_\phi,\ J_\psi$ to an extremal MP black hole of the same mass and angular momenta (therefore the corresponding horizon angular velocities are also equal). Nonetheless, they do not have equal horizon area or entropy. That is, \eqref{transj} and \eqref{cop} change the entropy, and as we will show below, enhance it.

To this end, we start from \eqref{2scharges} and evaluate the entropy and temperature of the ring in the collapsing limit:
\be\label{ring-collapse-entropy}
S^{ring}=\frac{\pi^2\tilde M\sqrt{2\tilde M}}{2G_5}\cdot \sqrt{\sigma}(1+\sqrt{1-4\sigma})  \,,\qquad T_{H}=\frac{1}{8\pi \sqrt{\tilde M}} \frac{\sqrt{2\sigma(1-4\sigma)}}{1+\sqrt{1-4\sigma}} (1-\nu)^2\to 0\,.
\ee
On the other hand one may compute the entropy of the extremal MP with the same  mass and angular momenta, i.e. evaluating \eqref{ext-MP-entropy} at \eqref{cop2} to obtain
\be\label{hole-ext-entropy}
S^{hole}=\frac{\pi^2\tilde M\sqrt{2\tilde M}}{2G_5}\cdot 2\sqrt{\sigma(1-\sigma)}\,.
\ee
As we see the hole has bigger entropy than the corresponding ring. In fact regardless of the value of angular momenta, one can show that $S^{ring}< S^{hole}\leq \sqrt3S^{ring}$. In other words, \eqref{transj} maps the system of black ring to a black hole of the same quantum numbers but with larger entropy, both at zero temperature. Therefore, the hole is a more stable configuration than the ring and one would expect quantum mechanical tunneling from the ring to the hole configuration.

We note that as the above analysis clearly indicates, it is not possible to approach to the vanishing horizon point $\lambda=2,\ \nu=1$ while keeping mass $M$ and $S/T$ finite; in this case, regardless of the value of $\sigma$ (even if we considered $\sigma\to 0$ limit),  $T/S\sim (1-\nu)^2\to 0$.  This is forced on us by the range of parameters and \eqref{nu-lambda-range}.

One may ask if taking the above collapsing limit and the near horizon limit commute. That is, do we get the same geometry if we took the $\lambda\to 2$ limit over the near horizon limit of generic extremal ring \eqref{dringNH}, and when we took the near horizon geometry of extremal MP?\footnote{Recall that as discussed above there is a coordinate transformation and a limit which involves cutting the horizon and completing it into a surface of topology $S^3$ and hence brings the ring geometry in the $\lambda\to 2,\ \nu\to 1$ to that of extremal MP black hole. This procedure changes the entropy and hence one would not expect the two limits commute. Nonetheless, it is worth seeing this explicitly at the level of metric too.} The latter has been presented in \eqref{nhmp} and the former may be obtained upon the parameter and coordinate transformations:
\be
\frac{2k}{2-\lambda}=\tilde k\equiv \sqrt{\frac{\tilde M}{8}}\,, \qquad x=-1+\frac{2-\lambda}{2}\tan^2\frac{\theta}{2}\,,\quad \theta\in [0,\pi-(2-\lambda)]\,,\qquad 2-\lambda\to 0\,,
\ee
yielding
\be\label{NHring-collapse}
ds^2=\tilde k^2\left[\frac{1+\cos^2\theta}{2}(-y^2dt^2+ \frac{dy^2}{y^2})+ \frac{1+\cos^2\theta}{2} d\theta^2+\frac{2\sin^2\theta}{1+\cos^2\theta}(d\varphi-y dt)^2+16d\psi^2 \right],
\ee
where $\varphi=\phi+3\psi$, $\varphi,\psi\in [0,2\pi]$ and $\theta\in[0,\pi]$. This geometry is basically the same as NHEK$\times S^1$ (where NHEK stands for near horizon extremal Kerr \cite{Guica:2008mu}). One may readily see that
the area of constant $y,\ t$ part of this geometry is equal to  $4\cdot2\pi \cdot 4\pi \tilde k^3$ which is the same as \eqref{horizon-area-extremal-ring}.

Metric \eqref{NHring-collapse} should be compared with $b=3a$ case in \eqref{nhmp}, where $\tilde k=a$. This is due to the fact that we have set $\nu=\lambda^2/4$ which corresponds to $\sigma=1/4$.
As we see the limits do not commute and this geometry is not the near horizon extremal MP \eqref{nhmp}. Moreover, as is seen from \eqref{ring-collapse-entropy} and \eqref{hole-ext-entropy},
entropies do not equal either: $S^{hole}=\sqrt3 S^{ring}$.

\subsection{EVH and near-EVH rings}

The analysis of previous subsection reveals that the only possibility of finding an EVH ring is around $\lambda=2,\ \nu=1$ point. To this end, let us consider
the following generic scaling
\be\label{EVH-ring-scaling}
\nu=1-\hat \nu \epsilon \,,\qquad \lambda=1+\nu-\hat\lambda \epsilon^{2(1+\alpha)}\,,\qquad k=\hat k \epsilon^{1+\alpha}\,,
\ee
where $\alpha\geq 0$ (coming from the range of parameters \eqref{nu-lambda-range}) and the scaling of $k$ is fixed by demanding having a finite mass $M$.
In the notation of previous subsection $\sigma\sim \epsilon^{2\alpha}$. Therefore, the $\alpha=0$ case reproduces a finite entropy black ring. To get a vanishing horizon limit we need to take $\alpha>0$. Moreover, for the $\alpha>0$ EVH case, as can be readily seen from \eqref{ring-collapse-entropy} and \eqref{hole-ext-entropy}, the hole and ring will have similar entropy. That is, in the near-EVH limit ring and hole cannot be distinguished by their entropy;\footnote{Note that the ``extremal ring'' case with strict $\nu=\lambda^2/4$ corresponds to $\sigma=1/4$ case and hence metric \eqref{dringNH} or \eqref{NHring-collapse} does not correspond to an EVH ring.} in this case,
$$
S_{BH}=\frac{\pi^2 \tilde M\sqrt{2\tilde M \hat\lambda}}{G_5 \hat\nu} \epsilon^{\alpha}\,,\qquad T_H\sim \epsilon^{2+\alpha}\,,
$$
and hence $T/S\sim \epsilon^2$ independently of the value of $\alpha$. This means that the above expansion does not allow non-extremal excitations of the ``EVH ring''. This is in contrast with the generic near-EVH MP black hole and the key difference between the two cases.\footnote{We comment that the above analysis has been carried out for \emph{balanced} rotating ring. As has been discussed in \cite{unbalanced-ring,Elvang:2007hs} it is possible to get a generic non-extremal MP black hole if one considered unbalanced ring. We will discuss this further in the discussion section.}

One may now take the near horizon, near-EVH ring. For the latter we use \eqref{EVH-ring-scaling} and choose $\alpha=1$. (This choice of $\alpha$ is made to match with parametrization used in section \ref{near-EVH-hole-section}.) In this case using transformations (\ref{transj}), the solution goes to an extremal MP black hole with near horizon metric in the form of (\ref{nhmp}). On the other hand as it is mentioned above, by the choice of $\alpha=1$, horizon area and the rotation parameter $a$ of MP black hole tend to zero; so the near horizon, near-EVH limit can be obtained by considering an infinitesimal value for the rotation parameter $a$. The result is
\be \label{nselfd}
ds^2=\sin^2\theta\,\bigg[-\frac{r^2}{4b^2} dt^2+b^2 \frac{dr^2}{4r^2}+{p^+}^2\,\big(d\phi+\frac{1}{2\,p^+\,b}\,r dt\big)^2\bigg]+b^2\,\big(\sin^2\theta\,d\theta^2+\cot^2\theta d\psi^2\big)\,,
\ee
in the above $t=\frac{\hat t}{b},\ \hat r=r$ and  $p^+=\sqrt{ab}$. This is a self-dual AdS$_3$ orbifold with radius $b$ \cite{deBoer:2010ac} with infinitesimal light-cone momentum $p^+/b$.

\section{Dual CFT descriptions of extremal rings and holes}\label{section-4}

We showed in the last section that around the $\lambda=2, \nu=1$ corner of the black ring parameter space, the ring geometry collapses into an extremal MP black hole with the same mass and angular momenta, while the hole generically has a larger entropy than the corresponding ring. The two entropies, however, match in the ``EVH'' limit of both ring and hole. In this section we argue how these facts can be seen from the proposed dual 2d CFT descriptions of the extremal black hole and extremal black ring, a la Kerr/CFT \cite{Guica:2008mu,Compere-review}, and for the near-EVH hole and ring from the EVH/CFT proposal \cite{SheikhJabbaria:2011gc}.

\subsection{Kerr/CFT description of extremal black ring and MP black hole}\label{Kerr-CFT-ring-hole}

\paragraph{Kerr/CFT description of extremal MP black hole.}  There are two chiral 2d CFT descriptions associated with generic extremal near horizon MP black hole \eqref{nhmp}, one associated with the ring direction $\psi$ and the other with the $\phi$ direction \cite{Lu:2008jk}.
The corresponding central charges may be computed using usual Kerr/CFT techniques \cite{Compere-review,Lu:2008jk}
\bea \label{mpcc}
c_{\phi}=\frac{3\,\pi}{2\,G_5} \left( a+b \right) ^{2}b=6J_\psi\,,\quad \qquad c_{\psi}=\frac{3\,\pi}{2\,G_5} \left( a+b \right) ^{2}a=6J_\phi\,.
\eea
The black hole corresponds to thermal states in these chiral 2d CFTs at temperatures
\be\label{FT-phi-psi}
T_\phi=\frac{1}{2\pi k_\phi}=\frac{1}{\pi}\sqrt{\frac{a}{b}}\,,\qquad T_{\psi}=\frac{1}{2\pi k_\psi}=\frac{1}{\pi}\sqrt{\frac{b}{a}}\,,
\ee
and the entropy is given by the Cardy formula
\be
S_{BH}=\frac{\pi^2}{3} c_\phi T_\phi=\frac{\pi^2}{3} c_\psi T_\psi\,.
\ee
\paragraph{Kerr/CFT description of extremal black ring.}
It is shown that  \cite{{Chen:2012yd}, cringholography} extremal black ring also admits Kerr/CFT description, that is the CFT description associated with geometry \eqref{dringNH}. Unlike the MP black hole case, however, there is no 2d CFT associated with the ring direction $\psi$ and we only have a single chiral 2d CFT description associated with the $\phi$ direction. This latter may also be seen from the form of the near horizon extremal ring metric \eqref{dringNH} where the geometry is of the form NHEK$\times S^1$ and the circle is the ring direction $\psi$.\footnote{The $\psi$ direction does not contribute to the entropy via Kerr/CFT. This fact has been noted previously in \cite{Ring-microstates}. The picture provided in these works is that the other ``chiral" sector of the presumed 2d CFT not appearing in the chiral CFT, is put to the ground state and is responsible for supporting the ring direction. This picture resembles, but not exactly the same as, what we advocate for the near-EVH ring in section \ref{EVH-CFT-section}.} The central charge of this 2d CFT is given by
\be \label{ccdrbr}
	\hat c_{\phi}={\frac {384\pi \,{\lambda}^{2}{k}^{3}}{G_5 \left( 2+\lambda \right)
 \left( 2-\lambda \right) ^{3}}}=12 J_\phi\,, 
\ee
which produces the entropy $\frac{A_H}{4G_5}$ (\emph{cf.} \eqref{horizon-area-extremal-ring}) for the CFT at temperature $T_{\phi}=\frac{1}{2\pi}$.

\paragraph{Extremal hole vs extremal ring.} As discussed, in the $\lambda\to 2$ limit we have a configuration of extremal black ring and an extremal black hole of the same mass and spin (this happens for $J_\psi=3J_\phi, G_5M^3_{phys.}=27\pi J^2_\phi/2$) but with the hole having a larger entropy (by a factor of $\sqrt3$) than the ring.
In this case the two $\phi$ and $\psi$ CFTs of the hole have central charges and temperatures as
\be\label{hole-CFTs}
\begin{split}
c_\phi=18J_\phi\,,\quad T_\phi=\frac{\sqrt3}{3\pi}\,;\qquad
c_\psi=6J_\phi\,,\quad T_\psi=\frac{\sqrt3}{\pi}\,;\qquad S^{hole}=2\pi\sqrt3 J_{\phi}\,,
\end{split}
\ee
while in the extremal ring case, the central charge and temperature of the dual chiral 2d CFT is
\be\label{ring-CFT}
\hat c_\phi=12 J_\phi \,,\qquad T_{\phi}=\frac{1}{2\pi}\,,\qquad S^{ring}=2\pi J_{\phi}\,.
\ee
As we see neither of the hole CFTs actually match that of the ring. This means that the ring and the hole are basically two different states in two distinct 2d chiral CFTs. It is interesting to note that the hole CFT has a larger central charge.

\subsection{EVH hole/ring, and their dual CFT description}\label{EVH-CFT-section}

As discussed in section \ref{section-2}, in the near horizon limit of an EVH MP black hole we find a pinching AdS$_3$ with radius $\ell_3=b$ and the effective 3d Newton constant $G_3=\frac{G_5}{\pi b^2}$.  Based on this AdS$_3$ factor the EVH/CFT proposal was stated in \cite{SheikhJabbaria:2011gc,deBoer:2011zt}: low energy excitations around an EVH black hole are described by a dual 2d CFT at the Brown-Hennueax central charge \cite{BH}.
Ignoring the pinching for the moment, the Brown-Hennueax central charge of the 2d CFT associated with the AdS$_3$ factor is
\be \label{extcharg}
c_{B.H.}=\frac{3\ell_3}{2G_3}=\frac{3\,\pi}{2\,G_5}\, b^3=c_{\phi}\,,
\ee
where $c_\phi$ is the expression for the Kerr/CFT central charge \eqref{mpcc} for the EVH case of $a=0$. In the EVH case the other central charge $c_\psi$ vanishes.

In the near-EVH case, with $a\sim \sigma\ll 1$ either of the ring and the hole  correspond to a specific state in the 2d CFT associated with the pinching AdS$_3$. One should note that, as discussed, one cannot have non-extremal excitations of the ring while as near horizon near-EVH geometry \eqref{near-EVH-MP}
clearly shows, an EVH MP black hole admits generic non-extremal excitation. Therefore, to compare the two cases with the same quantum numbers one should focus on the extremal excitations of the EVH MP black hole.
Both of these states have the same entropy, $S=2\pi b^3\sqrt\sigma/G_5$. However, as the near horizon geometries show, for the hole \eqref{near-EVH-MP} we are dealing with extremal BTZ and for the ring \eqref{nselfd} with a self-dual AdS$_3$ orbifold. So, the 2d CFT distinction between the hole and the ring
lies within the 2d CFT distinction of an extremal BTZ from a self-dual orbifold. As discussed in
\cite{Balasubramanian:2009bg}, the former corresponds to the thermal state $|T_L=0\rangle\bigotimes |T_R\rangle$, where $|T\rangle$ is a thermal state at temperature $T$, while the latter (the self-dual orbifold) corresponds to $|c/24 \rangle\bigotimes |T_R\rangle$, where $|c/24\rangle$ is the ground state of the 2d CFT on the plane and has energy $c/24$ above the vacuum of the 2d CFT on the cylinder.

To summarize this section, we discussed that
as we approach the EVH point one of the CFT's ($\psi$-CFT) of the Kerr/CFT duals of extremal MP hole (\emph{cf}. section  \ref{Kerr-CFT-ring-hole}) becomes singular  as its central charge goes to zero, while another possibility opens up: the other chiral sector of the $\phi$ direction becomes dynamically available for the hole case; giving rise to the 2d CFT proposed in the EVH/CFT. This latter, however, does not happen for the ring and we still remain with one chiral sector (of the $\phi$-CFT) which may now be viewed as the chiral CFT obtained through DLCQ of the 2d CFT appearing in the EVH/CFT.

One may then ask if it is possible to use EVH/CFT to understand the Kerr/CFT descriptions for generic extremal ring or hole, by viewing them as large excitations above the EVH geometry. As the corresponding near horizon geometries also indicate, the latter seems not possible. This is due to the fact that in taking the near-EVH limit we have already restricted ourselves to certain low energy excitations above the EVH point which excludes excitations with finite entropy, like those appearing in a generic extremal hole or ring.

\section{Discussion}\label{section-5}

In this work, with the goal of giving a 2d CFT dual distinction of extremal black holes from extremal black rings, we first made  a thorough review of 5d MP black holes and 5d black rings and their extremal and EVH limits. As we discussed there are regions of the parameter space of the ring, where it can be mapped to an extremal MP black hole. We hence focused on this region where there are dual CFT proposals for both the ring and the hole. As we discussed despite having the same mass and angular momenta, the ring and the hole do not generically have the same entropy; the geometry with larger entropy is the hole. In the dual CFT descriptions of the hole and the ring, this showed itself in the fact that the Kerr/CFT central charges (and temperatures) of the chiral 2d CFTs dual to the hole and the ring are different. That is, the ring and the hole are two thermal states in two completely distinct chiral 2d CFTs. Nonetheless, there are ``non-generic'' points in the overlapping region of hole/ring parameter space, the EVH region, where the ring and the hole of equal spins and angular momenta have equal entropy. For these cases, the hole and the ring correspond to two different states in the same 2d CFT, the one appearing in the EVH/CFT proposal. Of course, the chiral 2d CFT of the proposed Kerr/CFT in the near-EVH limit reduces to a chiral sector of this 2d CFT. In this way one can distinguish a near-EVH hole from a near-EVH ring in the dual EVH/CFT along the lines of \cite{Balasubramanian:2009bg}.

In this work we mainly focused on how one can distinguish hole from rings of the same quantum numbers. However, as we discussed the hole has a larger entropy than the ring. It is interesting to study if and how the ring to hole transition/tunneling can take place and how this will appear in the dual CFT pictures.
In the near-EVH case, the hole to ring transition corresponds to the DLCQ procedure \cite{Balasubramanian:2009bg} and the reverse, the ring to hole transition should appear as a ``thermalization in the ground state'', i.e. replacing the $|c/24\rangle$ ground  state with a thermal state at zero temperature. It is desirable to study this latter in more detail.

As mentioned in section \ref{ring-parameter-space-section}, the $\lambda=1+\nu$ line in the parameter space of the ring is a singular region. Nonetheless, in the $\lambda\to 1+\nu$ limit we find rings of finite mass and single spin with vanishing entropy and temperature while  $S/T$ remains finite. This region seems to be falling into our definition of EVH black holes. It is desirable to study this region of the ring parameter space in more detail and see if one can find dual CFT descriptions for. This problem is seemingly related to a more general class of black ring solutions, the \emph{unbalanced rings} where the centrifugal force on the ring is not balanced by its self-gravity. The unbalanced ring solutions form a four parameter family and one can recover the doubly rotating (balanced) ring we considered here as a spacial limit of them \cite{unbalanced-ring}. We expect that the $\lambda=1+\nu$ line should correspond to a particular single spin family of unbalanced
rings. A detailed analysis of this latter is postponed to upcoming publications.

\section*{Acknowledgement}

We would like to thank Roberto Emparan, Joan Sim\'on and Hossein Yavartanoo  for fruitful discussions and comments on the draft. AG and HG would like to thank IPM for hospitality while this project was completed. HG  would also like to thank Davood Mahdavian Yekta for useful discussions.
The work of AG and HG is supported by Ferdowsi University of Mashhad under the grant 3/22376 (1391/04/13).




\begin{thebibliography}{99}

\bibitem{HE}
S. W. Hawking and G. F. R. Ellis, ``The large scale structure of space-time,'' Cambridge
University Press, Cambridge, 1973.

  \bibitem{MP}
  R.~C.~Myers and M.~J.~Perry,
 ``Black Holes in Higher Dimensional Space-Times,''
  Annals Phys.\  {\bf 172}, 304 (1986).

  \bibitem{Emparan:2001wn}
  R.~Emparan and H.~S.~Reall,
  ``A Rotating black ring solution in five-dimensions,''
  Phys.\ Rev.\ Lett.\  {\bf 88}, 101101 (2002)
  [hep-th/0110260].

\bibitem{Pomeransky:2006bd}
  A.~A.~Pomeransky and R.~A.~Sen'kov,
  ``Black ring with two angular momenta,''
  hep-th/0612005.

  \bibitem{Emparan:2008eg}
  R.~Emparan and H.~S.~Reall,
  ``Black Holes in Higher Dimensions,''
  Living Rev.\ Rel.\  {\bf 11}, 6 (2008)
  [arXiv:0801.3471 [hep-th]].

\bibitem{Elvang:2007hs}
  H.~Elvang and M.~J.~Rodriguez,
  ``Bicycling Black Rings,''
  JHEP {\bf 0804} (2008) 045
 [arXiv:0712.2425 [hep-th]].


\bibitem{Lu:2008jk}
  H.~Lu, J.~Mei and C.~N.~Pope,
  ``Kerr-AdS/CFT Correspondence in Diverse Dimensions,''
  JHEP {\bf 0904}, 054 (2009)
  [arXiv:0811.2225 [hep-th]].


\bibitem{Chen:2012yd}
 B.~Chen and J.~-j.~Zhang,
  ``Holographic Descriptions of Black Rings,''
  JHEP {\bf 1211}, 022 (2012)
  [arXiv:1208.4413 [hep-th]].

\bibitem{D1D5P}
H.~Elvang, R.~Emparan,
``Black rings, supertubes, and a stringy resolution of black hole nonuniqueness,''
  JHEP {\bf 0311}, 035 (2003)
  [hep-th/0310008].

I.~Bena and P.~Kraus, ``Microscopic description of black rings in AdS / CFT,''
  JHEP {\bf 0412}, 070 (2004)
  [hep-th/0408186].

  \bibitem{SheikhJabbaria:2011gc}
  M.~M.~Sheikh-Jabbari and H.~Yavartanoo,
  ``EVH Black Holes, AdS3 Throats and EVH/CFT Proposal,''
  JHEP {\bf 1110}, 013 (2011)
  [arXiv:1107.5705 [hep-th]].


\bibitem{Guica:2008mu}
  M.~Guica, T.~Hartman, W.~Song and A.~Strominger,
  ``The Kerr/CFT Correspondence,''
  Phys.\ Rev.\ D {\bf 80}, 124008 (2009)
  [arXiv:0809.4266 [hep-th]].

\bibitem{Compere-review}
G.~Compere,
``The Kerr/CFT correspondence and its extensions: a comprehensive review,''
  Living Rev.\ Rel.\  {\bf 15}, 11 (2012)
  [arXiv:1203.3561 [hep-th]].


 \bibitem{Balasubramanian:2009bg}
  V.~Balasubramanian, J.~de Boer, M.~M.~Sheikh-Jabbari and J.~Simon,
  ``What is a chiral 2d CFT? And what does it have to do with extremal black holes?,''
  JHEP {\bf 1002}, 017 (2010)
  [arXiv:0906.3272 [hep-th]].

\bibitem{EVH-examples}
R.~Fareghbal, C.~N.~Gowdigere, A.~E.~Mosaffa and M.~M.~Sheikh-Jabbari,
``Nearing Extremal Intersecting Giants and New Decoupled Sectors in N = 4 SYM,''
  JHEP {\bf 0808}, 070 (2008)
  [arXiv:0801.4457 [hep-th]]; ``Nearing 11d Extremal Intersecting Giants and New Decoupled Sectors in D = 3,6 SCFT's,''
  Phys.\ Rev.\ D {\bf 81}, 046005 (2010)
  [arXiv:0805.0203 [hep-th]].

  H.~Yavartanoo,
``On heterotic black holes and EVH/CFT correspondence,''
  Eur.\ Phys.\ J.\ C {\bf 72}, 2256 (2012); ``EVH Black Hole Solutions With Higher Derivative Corrections,''
  Eur.\ Phys.\ J.\ C {\bf 72}, 1911 (2012)
  [arXiv:1301.4174 [hep-th]]; ``On EVH black hole solution in heterotic string theory,''
  Nucl.\ Phys.\ B {\bf 863}, 410 (2012)
  [arXiv:1212.3742 [hep-th]]; ``Five-dimensional heterotic black holes and its dual IR-CFT,''
  Eur.\ Phys.\ J.\ C {\bf 72} (2012) 2197 [arXiv:1301.3706 [physics.gen-ph]].

\bibitem{deBoer:2011zt}
 J.~de Boer, M.~Johnstone, M.~M.~Sheikh-Jabbari and J.~Simon,
 ``Emergent IR Dual 2d CFTs in Charged AdS5 Black Holes,''
  Phys.\ Rev.\ D {\bf 85}, 084039 (2012)
  [arXiv:1112.4664 [hep-th]].
\bibitem{Johnstone:2013eg}
M.~Johnstone, M.~M.~Sheikh-Jabbari, J.~Simon and H.~Yavartanoo,
``Near-Extremal Vanishing Horizon AdS5 Black Holes and Their CFT Duals,''
  JHEP {\bf 1304}, 045 (2013)
  [arXiv:1301.3387 [hep-th]].

\bibitem{Johnstone:2013ioa}
  M.~Johnstone, M.~M.~Sheikh-Jabbari, J.~Simon and H.~Yavartanoo,
  ``Extremal Black Holes and First Law of Thermodynamics,''
  arXiv:1305.3157 [hep-th].
\bibitem{deBoer:2010ac}
  J.~de Boer, M.~M.~Sheikh-Jabbari and J.~Simon,
 ``Near Horizon Limits of Massless BTZ and Their CFT Duals,''
  Class.\ Quant.\ Grav.\  {\bf 28}, 175012 (2011)
  [arXiv:1011.1897 [hep-th]].




  \bibitem{Hawking-Hartle}
S.~W.~Hawking, C.~J.~Hunter and M.~Taylor,
 ``Rotation and the AdS / CFT correspondence,''
  Phys.\ Rev.\ D {\bf 59}, 064005 (1999)
  [hep-th/9811056].

  \bibitem{Bardeen:1999px}
  J.~M.~Bardeen and G.~T.~Horowitz,
  ``The Extreme Kerr throat geometry: A Vacuum analog of AdS(2) x S**2,''
  Phys.\ Rev.\ D {\bf 60}, 104030 (1999)
  [hep-th/9905099].

  \bibitem{Matsuo:2010ut}
  Y.~Matsuo and T.~Nishioka,
  ``New Near Horizon Limit in Kerr/CFT,''
  JHEP {\bf 1012}, 073 (2010)
  [arXiv:1010.4549 [hep-th]].

  \bibitem{Kunduri:2013gce}
  H.~K.~Kunduri and J.~Lucietti,
  ``Classification of near-horizon geometries of extremal black holes,''
  arXiv:1306.2517 [hep-th].


\bibitem{cringholography}
Ahmad Ghodsi, Hanif Golchin, M.M. Sheikh-Jabbari, ``work in progress''

 \bibitem{Matsumoto:2012cg}
  M.~Matsumoto, H.~Yoshino and H.~Kodama,
 ``Time evolution of a thin black ring via Hawking radiation,''
  arXiv:1205.5715 [gr-qc].

\bibitem{unbalanced-ring}
Y.~Morisawa, S.~Tomizawa and Y.~Yasui,
``Boundary Value Problem for Black Rings,''
  Phys.\ Rev.\ D {\bf 77}, 064019 (2008)
  [arXiv:0710.4600 [hep-th]].

  Y.~Chen, K.~Hong and E.~Teo,
``Unbalanced Pomeransky-Sen'kov black ring,''
  Phys.\ Rev.\ D {\bf 84}, 084030 (2011)
  [arXiv:1108.1849 [hep-th]].



\bibitem{Ring-microstates}
J.~J.~Blanco-Pillado, R.~Emparan and A.~Iglesias,
``Fundamental Plasmid Strings and Black Rings,''
  JHEP {\bf 0801} (2008) 014
  [arXiv:0712.0611 [hep-th]].

R.~Emparan,
  ``Exact Microscopic Entropy of Non-Supersymmetric Extremal Black Rings,''
  Class.\ Quant.\ Grav.\  {\bf 25}, 175005 (2008)
  [arXiv:0803.1801 [hep-th]].

\bibitem{BH}J. D. Brown and M. Henneaux, {\it Central Charges in the Canonical Realization of
 Asymptotic Symmetries: An Example from Three-Dimensional Gravity},
 Commun. Math. Phys. 104 (1986) 207�226;

\end{thebibliography}
 \end{document}